\documentclass[runningheads]{llncs}

\usepackage[T1]{fontenc}
\usepackage{amssymb}
\usepackage[binary-units=true]{siunitx}
\usepackage{graphicx}
\usepackage{glossaries}
\usepackage{xcolor}
\usepackage{booktabs}
\usepackage[shortlabels]{enumitem}
\usepackage{subcaption}
\usepackage{tumcolor}
\usepackage{bbding}
\usepackage[colorlinks=true]{hyperref}

\def\yes{\textcolor{TUMDarkerGreen}{\large\checkmark}}
\def\maybe{\textcolor{TUMOrange}{\Large$\circ$}} % $\mathbit
\def\no{\textcolor{TUMRed}{\Large\texttimes}}

\DeclareSIUnit[per-mode=symbol]\bps{\bit\per\second}
\DeclareSIUnit[per-mode=symbol]\kbps{\kilo\bps}
\DeclareSIUnit[per-mode=symbol]\Mbps{\mega\bps}
\DeclareSIUnit[per-mode=symbol]\Gbps{\giga\bps}
\DeclareSIUnit[per-mode=symbol]\nanosec{\nano\second}
\DeclareSIUnit[per-mode=symbol]\packet{packet}
\DeclareSIUnit[per-mode=symbol]\packetps{\packet\per\second}
\DeclareSIUnit\microsec{\SIUnitSymbolMicro s}
\DeclareSIUnit\byte{B}
\DeclareSIUnit\bit{bit}
\DeclareSIUnit\terabyte{TB}

\usepackage[backend=biber,
  doi=true,
  isbn=false,
  url=true]{biblatex}
\begin{document}

\title{Threshold Signatures for Central Bank Digital Currencies}

\author{
  Mostafa Abdelrahman\inst{1,2}\and
  Filip Rezabek\inst{1}\orcidID{0000-0002-9090-5633}\and\\
  Lars Hupel\inst{1,2}\Envelope\orcidID{0000-0002-8442-856X}\and
  Kilian Glas\inst{1}\and
  Georg Carle\inst{1}
}
\authorrunning{M. Abdelrahman et al.}
\institute{Technische Universität München, Munich, Germany \and Giesecke+Devrient, Munich, Germany\\
\email{\Envelope\ lars.hupel@tum.de}
}

\maketitle

\begin{abstract}
Digital signatures are crucial for securing Central Bank Digital Currencies (CBDCs) transactions.
Like most forms of digital currencies, CBDC solutions rely on signatures for transaction authenticity and integrity, leading to major issues in the case of private key compromise.
Our work explores threshold signature schemes (TSSs) in the context of CBDCs.
TSSs allow distributed key management and signing, reducing the risk of a compromised key.
We analyze CBDC-specific requirements, considering the applicability of TSSs, and use Filia CBDC solution as a base for a detailed evaluation.
As most of the current solutions rely on ECDSA for compatibility, we focus on ECDSA-based TSSs and their supporting libraries.
Our performance evaluation measured the computational and communication complexity across key processes, as well as the throughput and latency of end-to-end transactions.
The results confirm that TSS can enhance the security of CBDC implementations while maintaining acceptable performance for real-world deployments.

\end{abstract}

\newacronym{nic}{NIC}{Network Interface Card}
\newacronym{iiot}{IIoT}{Industrial Internet of Things}
\newacronym{iot}{IoT}{Internet of Things}
\newacronym{opc}{OPC}{Open Platform Communications}
\newacronym{opcua}{OPC UA}{Open Platform Communications Unified Architecture}
\newacronym{can}{CAN}{Controller Area Network}
\newacronym{lin}{LIN}{Local Interconnect Network}
\newacronym{tas}{TAS}{Time-Aware Shaper}
\newacronym{cbs}{CBS}{Credit-Based Shaper}
\newacronym{tsn}{TSN}{Time Sensitive Networking}
\newacronym{taprio}{TAPRIO}{Time Aware Priority Shaper}
\newacronym{etf}{ETF}{Earliest Time First}
\newacronym{mqprio}{MQPRIO}{Multiqueue Priority Qdisc}
\newacronym{cots}{COTS}{Commercial off-the-Shelf}
\newacronym{rtt}{RTT}{Round Trip Time}
\newacronym{ua}{UA}{Unified Architecture}
% \newacronym{opcua}{OPC UA}{OPC Unified Architecture}
\newacronym{e2e}{E2E}{End-to-End}
\newacronym{p2p}{P2P}{Peer-to-Peer}
\newacronym{ptp}{PTP}{Precision Time Protocol}
\newacronym{lidar}{LiDAR}{Light Detection and Ranging}
\newacronym{gptp}{gPTP}{generic Precision Time Protocol}
\newacronym{phc}{PHC}{PTP Hardware Clock}
\newacronym{jbod}{JBOD}{Just a Bunch of Devices}
\newacronym{avb}{AVB}{Audio Video Bridging}
\newacronym{gm}{GM}{Grandmaster Clock}
\newacronym{pubsub}{PubSub}{Publish–Subscribe}
\newacronym{tc}{tc}{traffic control}
\newacronym[plural=TCLs,firstplural=traffic classes (TCLs)]{tcl}{TCL}{Traffic Class}
\newacronym{sr}{SR}{Stream Reservation}
\newacronym{qdisc}{qdisc}{queuing discipline}
\newacronym{qos}{QoS}{Quality of Service}
\newacronym{ecdf}{ECDF}{Empirical Cumulative Distribution Function}
\newacronym{uadp}{UADP}{Unified Architecture Datagram Protocol}
\newacronym{be}{BE}{Best Effort}
\newacronym{kpi}{KPI}{Key Performance Indicator}
\newacronym{pcp}{PCP}{Priority Code Point}
\newacronym{candc}{C\&C}{Command \& Control}
\newacronym{rtaw}{RTaW}{Real-Time at Work}
\newacronym{skb}{SKB}{Socket Buffer}
\newacronym{ivn}{IVN}{Intra-Vehicular Network}
\newacronym{zgw}{ZGW}{Zonal Gateway}
\newacronym{vcc}{VCC}{Vehicular Control Computer}
\newacronym{oc}{OC}{Ordinary Clock}
\newacronym[plural=TRCLs,firstplural=Transparent Clocks (TRCLs)]{tclo}{TRCL}{Transparent Clock}
\newacronym{bc}{BC}{Boundary Clock}
\newacronym{pfifo}{PFIFO}{Packet Limited First In, First Out queue}
\newacronym{sut}{SUT}{System Under Test}
\newacronym{adas}{ADAS}{Advanced Driver Assistance Systems}
\newacronym{phy}{PHY}{Physical Layer}
\newacronym{udp}{UDP}{User Datagram Protocol}
\newacronym{bmca}{BMCA}{Best Master Clock Algorithm}
\newacronym{tcp}{TCP}{Transmission Control Protocol}
\newacronym{os}{OS}{Operating System}
\newacronym{irq}{IRQ}{Interrupt Request}
\newacronym{cpu}{CPU}{Central Processing Unit}
\newacronym{smp}{SMP}{Symmetrical Multiprocessing}
\newacronym{smt}{SMT}{Simultaneous Multi-Threading}
\newacronym{rt}{RT}{Real-Time}
\newacronym{hw}{HW}{hardware}
\newacronym{sw}{SW}{software}
\newacronym{dvfs}{DVFS}{Dynamic Voltage and Frequency Scaling}
\newacronym{vlan}{VLAN}{Virtual LAN}
\newacronym{frer}{FRER}{Frame Replication and Elimination for Reliability}
\newacronym{kc}{KC}{Key Contribution}
\newacronym{pcap}{PCAP}{Packet Capture}
\newacronym{utc}{UTC}{Coordinated Universal Time}
\newacronym{tai}{TAI}{International Atomic Time}
\newacronym{sdn}{SDN}{Software Defined Networking}
\newacronym{iic}{IIC}{Industrial Internet Consortium}
\newacronym{tdma}{TDMA}{Time Division Multiple Access}
\newacronym{ao}{AO}{Automatic Overclocking}
\newacronym{txtime}{TxTime}{transmission time}
\newacronym{macsec}{MACsec}{Media Access Control Security}
\newacronym{omnet}{OMNeT++}{Objective Modular Network Testbed in C++}
\newacronym{sa}{SA}{Security Association}
\newacronym{mka}{MKA}{MACsec Key Agreement}
\newacronym{eap}{EAP}{802.1x Extensible Authentication Protocol}
\newacronym{edgar}{EDGAR}{Excellent Driving GARching}
\newacronym{hpc}{HPC}{High Performance Computer}
\newacronym{lpc}{LPC}{Low Performance Computer}
\newacronym{engine}{EnGINE}{Environment for Generic In-vehicular Networking Experiments}
\newacronym{sc}{SC}{Secure Channel}
\newacronym{dpdk}{DPDK}{Data Plane Development Kit}
\newacronym{sss}{SSS}{Shamir's secret sharing}
\newacronym{fpga}{FPGA}{Field Programmable Gate Array}
\newacronym{fqcodel}{FQ\_CoDel}{Fair Queuing with Controlled Delay}
\newacronym{mqtt}{MQTT}{Message Queuing Telemetry Transport}
\newacronym{mac}{MAC}{Media Access Control}
\newacronym{ip}{IP}{Internet Protocol}
\newacronym{ws}{WS}{window size}
\newacronym{ram}{RAM}{Random-Access Memory}
\newacronym{is}{IS}{Interframe Spacing}
\newacronym{ovs}{OvS}{Open vSwitch}
\newacronym{nc}{NC}{Network Calculus}
\newacronym{foi}{FOI}{Flow of Interest}
\newacronym{god}{GOD}{Guaranteed Output Delivery}
\newacronym{pp}{PP}{Payment Processor}
\newacronym{tfa}{TFA}{Total Flow Analysis}
\newacronym{gplusd}{G+D}{Giesecke+Devrient}
\newacronym[plural=CBDCs,longplural=Central Bank Digital Currencies]{cbdc}{CBDC}{Central Bank Digital Currency}
\newacronym{fsp}{FSP}{Financial Service Provider}
\newacronym{fips}{FIPS}{Federal Information Processing Standard}
\newacronym{dlp}{DLP}{Discrete Logarithm Problem}
\newacronym{dsa}{DSA}{Digital Signature Algorithm}
\newacronym{ecdsa}{ECDSA}{Elliptic Curve Digital Signature Algorithm}
\newacronym{eddsa}{EdDSA}{Edwards-curve DSA}
\newacronym{dkg}{DKG}{Distributed Key Generation}
\newacronym{zkp}{ZKP}{Zero-Knowledge Proof}
\newacronym{ppt}{PPT}{Probabilistic Polynomial Time}
\newacronym{kmn}{KMN}{Key Management Network}
\newacronym{poc}{PoC}{Proof-of-Concept}
\newacronym{methoda}{METHODA}{Multilayer Environment and Toolchain for Hollistic NetwOrk Design and Analysis}
\newacronym{mpc}{MPC}{Multiparty Computation}
\newacronym{ot}{OT}{Oblivious Transfer}
\newacronym{uc}{UC}{Universal Composability}
\newacronym{vCPUs}{vCPUs}{virtual CPUs}
\newacronym{tps}{TPS}{Transactions Per Second}
\newacronym{utxo}{UTXO}{unspent Transaction Output}
\newacronym{sota}{SotA}{State-of-the-Art}
\newacronym{tss}{TSS}{Threshold Signatures Scheme}
\newacronym{ucs}{UC}{use case}
\newacronym{pki}{PKI}{Public Key Infrastructure}
\newacronym{ddh}{DDH}{Decisional Diffie–Hellman}

% \newacronym{e2e}{E2E}{End-to-End}

\newcommand{\encircled}[2][0.8mm]{%
    \raisebox{.5pt}{%
        \textcircled{%
            \raisebox{0.35pt}{%
                \kern #1
                \scalebox{0.70}{#2}
            }%
        }%
    }%
}

\definecolor{ourgreen}{rgb}{0.00,0.49,0.19}
\definecolor{ourred}{rgb}{0.77,0.03,0.11}
\definecolor{ourorange}{rgb}{0.89,0.45,0.13}
\definecolor{ourgrey}{rgb}{0.60,0.60,0.60}
\def\yes{\textcolor{ourgreen}{\large\checkmark}}
\def\maybe{\textcolor{ourorange}{\Large$\circ$}} % $\mathbit
\def\no{\textcolor{ourred}{\Large\texttimes}}
\def\unknown{\textcolor{ourgrey}{\encircled[1mm]{?}}}
\section{Introduction}
\label{sec:introduction}
\makeatletter
\renewcommand\@makefnmark{}
\makeatother
Digital signatures are essential to ensure the authenticity and integrity of online data. 
They provide a way to verify the origin and integrity of a message, ensuring that the message has not been tampered with and that it indeed comes from the claimed sender. 
The system's security collapses if the private key $sk$ is compromised, as an attacker could forge signatures.
Conversely, if the $sk$ is lost or destroyed, valid signatures cannot be created, leading to availability loss. This overall leads to a single point of failure. \footnotetext[1]{Mostafa Abdelrahman and Filip Rezabek contributed equally to this paper.}

\glspl{tss} address the vulnerabilities associated with a single point of failure in private key management. 
The \gls{tss} distributes the signing authority among multiple parties, requiring a subset (or threshold $t$) of these parties to collaborate to produce a valid signature. 
This trust distribution reduces the risk of key compromise and increases the resilience of the cryptographic system. 
By eliminating a single point of failure, \glspl{tss} are particularly advantageous in high-stakes applications such as financial transactions and secure communications.

A key application of \glspl{tss} is within blockchain and digital currencies, which require robust security measures due to the significant value they protect~\cite{erinle2025sokdesignvulnerabilitiessecurity}. 
This is particularly relevant in the context of \glspl{cbdc}, which are digital representations of a nation's currency issued and regulated by the central bank.
\gls{cbdc} introduces the role of a \gls{fsp}, which process users' payments and ensure funds availability. 
In that case, the \gls{fsp} manages the private key of users as custodial key management. 
Naturally, a common design pattern uses online wallets distributed among various \glspl{fsp}.
However, this puts a lot of trust in the \glspl{fsp} to create and manage hosted wallets.
Besides, the \glspl{fsp} do not trust each other. 
Therefore, \gls{tss} is a viable solution as the threshold number of nodes within a \gls{fsp} have to collaborate, increasing the resilience of the \gls{fsp} system.
In general, we examine the integration of \gls{tss} in the context of \gls{cbdc} deployments.
For the scope of this paper, we select the Filia implementation \cite{gi-de-whitepaper}, an off-chain \gls{cbdc} solution developed by \gls{gplusd}. 
Filia follows a similar design with a focus on \glspl{fsp} for custodial key management, allowing the generalization of our findings to other \gls{cbdc} implementations with similar architecture.

Filia offers hardware-based (offline) and hosted (online) wallets. 
Applying threshold signatures to hardware wallets would be cumbersome for day-to-day transactions since it would require the user to juggle multiple devices.
Filia’s online wallets currently only support \gls{ecdsa}.
For that reason, we focus on \gls{sota} \gls{ecdsa} \gls{tss} as they apply to online wallets.
The goal is to develop a threshold version of this key management system to enhance security by preventing single-point failure. 

\paragraph{Contributions}
We provide these key contributions: 

\begin{itemize}
    \item Analysis of \gls{sota} \gls{tss} based on \gls{ecdsa};
    \item Protocol design fitting the \gls{cbdc} use cases;
    \item Theoretical and empirical evaluation of the protocol.
\end{itemize}
\section{Background}
\label{sec:background}
This section provides an overview of the CGGMP21 \gls{ecdsa} \gls{tss}, Filia as the selected case study \gls{cbdc}, and \gls{engine}/\gls{methoda} framework used for evaluation. 

\paragraph{CGGMP21}
The CGGMP21 protocol is a \gls{sota} \(n\)-out-of-\(n\) threshold \gls{ecdsa} scheme \cite{cryptoeprint:2021/060}. 
It provides features fitting the context of \gls{cbdc} applications.
First, it relies on Paillier encryption for secure multiplication of secret shares and utilizes \glspl{zkp} to detect and attribute deviations from the expected behavior \cite{10.1145/22145.22178,10.5555/1756123.1756146}.
Next, it supports pre-signing, identifiable aborts, security against adaptive adversaries, and universal composability under strong RSA, DDH, and Paillier security assumptions.
It provides proactive security with periodic key refreshes and ensures that compromised shares within an epoch do not affect the system's integrity.
It offers efficiency trade-offs: a 3-round presign protocol with \(O(n^2)\) identification cost or a 6-round presign protocol with \(O(n)\) identification cost.
The protocol can be extended to general \(t\)-out-of-\(n\) threshold case using \gls{sss} and converting Shamir secret shares into additive shares \cite{10.1145/359168.359176,10.1007/978-3-540-30576-7_19}.

\begin{figure}[t]
    \begin{subfigure}[t]{.43\linewidth}
    \includegraphics[width=1\linewidth]{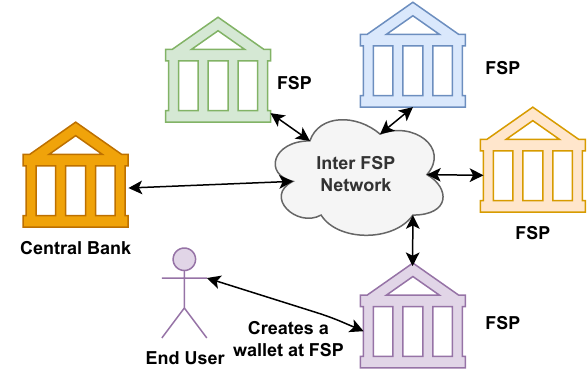}
    \caption{High-Level Architecture \cite{gi-de-whitepaper}}
    \label{fig:filia}
    \end{subfigure}
    \hfill
    \begin{subfigure}[t]{.55\linewidth}
        \includegraphics[width=1\linewidth]{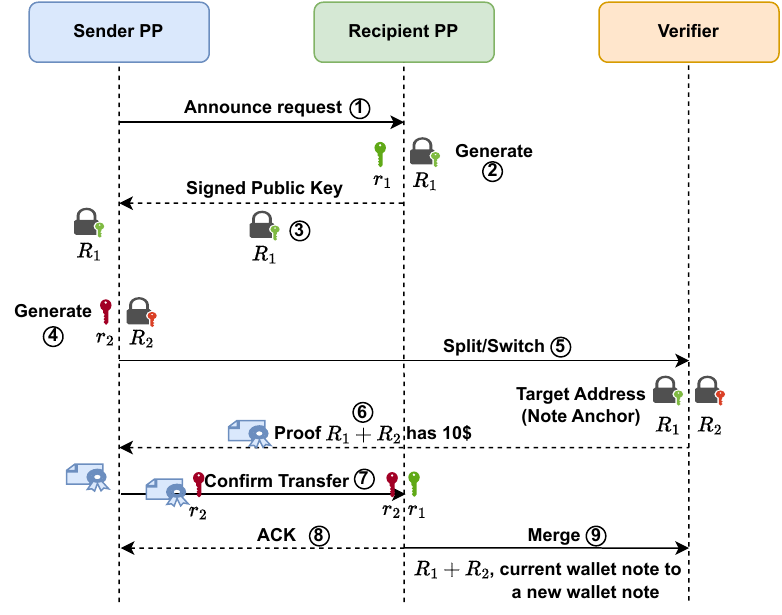}
         \caption{Filia Protocol: Remote Transaction. $r$ corresponds to private key and $R$ to public key. }
     \label{fig:transaction}
    \end{subfigure}
    \caption{Filia Architecture and Protocol Overview. Figures outline the involved parties and their form of interaction. Color differentiates between the parties.}
\end{figure}

\paragraph{Filia}
Figure~\ref{fig:filia} provides a high-level overview of the Filia system. 
The central bank is responsible for the core infrastructure. 
At the same time, private sector intermediaries, known as \glspl{fsp}, offer end-user wallets and develop services based on the technology and regulatory framework established by the central bank. 
Each \gls{fsp} maintains a wallet at the central bank, where the requested liquidity is transferred. 
Other \gls{cbdc} solutions with similar architecture exist. 

\paragraph{\acrshort{engine}/\acrshort{methoda}}
The \gls{engine}/\gls{methoda} framework \cite{Rezabek2022EngineJNSM,rezabek2023preprint}, initially developed for \gls{tsn} emulations, provides a robust and scalable environment designed to support reproducible and modular experiments in distributed cryptographic operations.
Built with Ansible-based orchestration, \gls{methoda} allows for flexible deployment and management of complex setups, making it well-suited for evaluating cryptographic protocols in distributed systems.
\section{Related Work}
\label{sec:relatedwork}
Existing research on \gls{cbdc} system design highlights different models, including two-tier architectures where central banks issue digital currency but rely on intermediaries for distribution \cite{CBDC_Design_2024} which Filia follows. 
However, security remains a critical challenge in \gls{cbdc} implementations, particularly in mitigating operational and cyber risks \cite{CBDC_Risk_2023}. 
This paper addresses these concerns by exploring \gls{tss} as a means to enhance key management security and reduce the risk of private key compromise. 

There is limited work in applying \glspl{tss} in the context of \gls{cbdc}, with existing research focusing primarily on distributed ledger-based implementations. 
PEReDi \cite{cryptoeprint:2022/974} employs \gls{tss} in a scenario where maintainers, including banks, financial institutions, and regulatory offices, each hold a share. 
This setup allows for selective transaction tracing and privacy revocation in cases of suspicious activities. 
IBM's \gls{cbdc} framework \cite{cryptoeprint:2023/1717} incorporates \gls{tss} within its new endorser model, where \glspl{tss} enhance security and accountability by requiring a collective agreement from multiple endorsers for transaction validation.
Overall, the framework focuses on hardening the token creation by the \gls{cbdc}, minimizing the likelihood of the creation of unauthorized tokens. 
On the other hand, our threat model aims to protect against theft or unauthorized use of tokens. 

Numerous initiatives within the cryptocurrency industry, many of which are part of the MPC Alliance \cite{mpc-a}, have adopted \glspl{tss} to secure digital assets. 
This distributed approach has enabled secure ``Wallet as a Service'' models~\cite{cryptoeprint:2022/974,dfns2024cggmp21,fireblocks2024}. 
These platforms enhance security by distributing cryptographic operations across multiple parties, ensuring robust protection even when certain parties may be compromised.
\section{Analysis \& Protocol Design}

The main \gls{cbdc} use case analyzed in this work considers the \gls{fsp} as the custodian of users' wallet private keys.
In this model, the \gls{fsp} securely manages private keys on behalf of multiple users and an entrusted central ledger 
maintains a record of public keys to ensure the legitimacy of the money.
This section focuses on the private key lifecycle and its distribution relying on the \gls{tss}.
First, we provide an overview of the Filia platform and possible solutions on how \gls{tss} can be integrated into it. 
We define general requirements relevant to the context of \gls{cbdc} along with a detailed security model.
We argue for selecting a suitable \gls{tss} based on the requirements. 

\subsection{Filia Protocol}
Filia relies on a modified \gls{utxo}-based model used to represent a digital currency. 
Each \gls{utxo}, referred to here as a \emph{note}, consists of a public-private key pair \((r, R)\) derived from the secp256r1/P-256 curve and a denomination \((v)\) indicating its monetary value. 
The private component \(r\) is securely managed within the hosted wallet by the \emph{\gls{pp}}, the main component of the \gls{fsp} responsible for wallet management and transaction handling. 
Meanwhile, the public component \(R\) is registered and tracked by a Verifier, a central bank infrastructure tasked with preventing double spending and ensuring system integrity.

Transactions in this system follow specific rules. 
Certain transaction types, such as currency issuance or destruction, are restricted to the central bank, while old transaction outputs are periodically garbage collected to improve system efficiency. 
Transaction commands involve consuming one or more existing notes and creating new ones, with the available commands being \textbf{Create}, \textbf{Split}, \textbf{Merge}, \textbf{Switch}, and \textbf{Destroy}. 
Each transaction is validated by the \textit{Verifier}, which ensures the correctness of commands and prevents double-spending. 
Notes are registered in the system only if they are unique, and commands are rejected if any input notes have already been consumed in a prior transaction.

The Filia system differentiates between two types of transfers: \textbf{local transfers} (wallets within the same \gls{fsp}) and \textbf{remote transfers} (wallets at different \gls{fsp}). 
While local transfers follow a simplified protocol, remote transfers require additional security measures due to the lack of trust between different \gls{fsp}. 
A non-repudiation protocol addresses this, ensuring that the sending \gls{pp} can provide proof of the money sent.
Figure~\ref{fig:transaction} illustrates the steps involved in a remote transfer under the Filia protocol:
\begin{enumerate}
    \item \textbf{Anounce Transfer:} The recipient \gls{pp} receives a announcement request from the sender \gls{pp} with value \((v)\) \encircled{1}. The recipient \gls{pp} generates a key pair \( (r_1, R_1) \) \encircled{2} and sends the public key \( R_1 \) back to the sender \gls{pp} \encircled{3}.
    \item \textbf{Register New Public Key:} The sender \gls{pp} generates a second key pair \( (r_2, R_2) \) \encircled{4} and sends a request to the verifier to confirm the transfer and register the value \((v)\) under \( R_1 + R_2 \) \encircled{5}.
    \item \textbf{Store Verifier Proof \encircled{6}:} The sender \gls{pp} stores a proof provided by the Verifier, demonstrating that the value \((v)\) has been transferred to the correct address \( R_1 + R_2 \) at the Verifier.
    \item \textbf{Confirm Transfer:} The sender \gls{pp} sends the private key \( r_2 \) to the recipient \gls{pp} \encircled{7}, allowing the recipient to access the funds and the recipient \gls{pp} replies with an acknowledgment \encircled{8}. 
\end{enumerate}

\noindent
Eventually, only the recipient \gls{pp} knows \( r_1 + r_2\) and can use the note. 
The recipient \gls{pp} can switch the funds to a new key or merge the funds to a previously available note \encircled{9}.

This process ensures that the recipient cannot access the transferred funds until the sender has securely stored the proof of transfer. Additionally, because the sender does not know the private key \( r_1 \), they cannot reclaim the money after it has been sent.

Should the sender PP not disclose \( r_2 \), the value \( v \) will be locked; however, it will also be unable to prove correct transfer execution.

\subsection{Possible Solutions}
Continuing with the protocol overview, we explore two potential solutions for implementing threshold key management within the Filia system at the \gls{fsp} level. 
We outline and analyze each solution, including its unique pros and cons.

\subsubsection{Solution 1: Payment Processor-Based Threshold Key Management}
In this approach, each \gls{fsp} deploys \( n \) \glspl{pp}, with each \gls{pp} holding a share of the private key. 
Specifically, each \gls{pp} holds a new note in the form \((v, r_1)\), \((v, r_2)\), ..., \((v, r_n)\). 
Here, the \gls{tss} is directly implemented within the payment processing system, so when a transaction requires signing, a predefined number of \gls{pp} collaborate to generate the signature.

\subsubsection{Solution 2: Separate \gls{kmn}}
This solution establishes a dedicated \gls{kmn} within the \gls{fsp} infrastructure, where a set of \( n \) specialized nodes (key management nodes) are responsible for generating and storing key shares, and providing signatures. 
\gls{pp} sends requests to the \gls{kmn} for signatures or key generation as needed. 
Each node in the \gls{kmn} holds a key share \( r_1, r_2, \ldots, r_n \), and assigns a specific identifier (\text{uuid}) to the shares, which is sent to the \gls{pp}. 
Consequently, the \gls{pp} holds a \texttt{note} in the form \((v, \text{uuid})\) and uses that \text{uuid} to request a signature from the \gls{kmn}.

\subsubsection{Discussion}
When considering the complexity, \textbf{Solution 1} tightly integrates key management into the payment processing infrastructure, requiring each \gls{pp} to maintain its database of key shares. 
This necessitates synchronization across multiple \glspl{pp}, significantly increasing the operational overhead, particularly in achieving consensus between \glspl{pp}. 
In contrast, \textbf{Solution 2} isolates key management in a dedicated network, simplifying the responsibilities of \glspl{pp}. 
Although this reduces complexity within the \glspl{pp}, it adds a layer of infrastructure that requires careful management, such as ensuring the availability and reliability of the \gls{kmn} nodes.

Regarding latency, \textbf{Solution 1} has an advantage due to the proximity of all operations within the \gls{fsp} infrastructure.
On the other hand, \textbf{Solution 2} introduces communication overhead as \glspl{pp} need to interact with the separate key management nodes. 
This added step can lead to higher latency, particularly in high-frequency transaction environments.

The modular design of \textbf{Solution 2} ensures that the \gls{tss} can be updated or replaced without requiring significant changes to the \gls{pp} codebase. 
The system can also take advantage of batching and parallelization to reduce delays and improve efficiency without disrupting other \gls{pp} tasks. 
By decoupling key management from the payment processing system, \gls{kmn} enhances security, as keys remain secure within the storage of dedicated nodes, even if \gls{pp} is compromised.
 
The \gls{kmn} is selected for its modular design, enhanced security, and ease of updates. 
Decoupling key management from payment processing eliminates the need for consensus mechanisms required in \textbf{Solution 1}, reducing the implementation complexity of \gls{poc}.
Furthermore, \gls{kmn} enables efficient batch processing and parallelization, ensuring scalability without disrupting payment processing tasks, making it a robust choice for the Filia system.

\subsection{System Requirements}
We outline several key requirements for extending the Filia protocol by \gls{tss} and considering \textbf{Solution 2}. 
They must easily migrate Filia wallets and transaction signing to the \gls{tss} system.
Notably, these requirements should be generalizable to designs similar to \gls{cbdc} as Filia.

\textbf{R1 - Transparent Support for Standard and Threshold Wallets:}
The system must conceal the type of wallet in use, whether standard or threshold, from external observers, including other \glspl{fsp}. 
This ensures flexibility within the Filia system, enabling different security levels based on factors such as the wallet's balance.

\textbf{R2 - Efficient \gls{dkg}:} Due to multiple generation of key during transactions in Filia, the \gls{dkg} process must be optimized for high efficiency. 

\textbf{R3 - High Signature Throughput:}
The signing process must achieve high throughput to handle the multiple signatures required per transaction and efficiently load high user loads. 
For instance, during a remote transfer, the process involves splitting the sender's wallet and merging the transferred value into the recipient's wallet (as shown in Figure~\ref{fig:transaction}, Steps \encircled{5} and \encircled{9}). 
These operations require multiple signatures, so the system must be capable of processing these quickly. 

\textbf{R4 - Secure Key Reconstruction:}  
The system must enable the secure reconstruction of short-lived keys when required. 
Specifically, when the sender's \gls{pp} generates a threshold key pair \( (r_2, R_2) \), it must securely reconstruct or export \( r_2 \) during the process of transmitting the full key to the recipient's \gls{pp} (e.g., Figure~\ref{fig:transaction}, Step \encircled{7}).

\textbf{R5 - Key Update/Key Addition:}  
The system must facilitate the secure addition of new key parts to existing key shares. 
For instance, as illustrated in Figure~\ref{fig:transaction}, the sender \gls{pp} must send \( r_2 \), which is a full key, to the recipient \gls{pp}. 
The recipient \gls{pp} then combines \( r_2 \) with their threshold key \( r_1 \). 
The result is that each node holds a share of the combined key \( r_1 + r_2 \).

\textbf{R6 - Compatibility with Existing Infrastructure:}
The \gls{tss} must be fully compatible with \gls{ecdsa}, in specific secp256r1/P-256 curve, to align with the existing infrastructure, particularly for hardware wallets that rely on smart card technology that use \gls{ecdsa}. 
This requirement ensures that the integration of \glspl{tss} does not disrupt existing systems and maintains consistent security across all wallet types.

\textbf{R7 - Security:}
The \gls{tss} should be secure with minimum assumptions and should be secure even when combined with other systems/protocols.

\subsection{Security System Model}
We define the environment and conditions under which the protocol is expected to operate.

\subsubsection{Adversary Model}
In our system, we assume the presence of an active, malicious adversary. 
This adversary can arbitrarily deviate from the protocol, aiming to disrupt the system or gain unauthorized access to information.

\textbf{Threshold Structure:}
We operate under the assumption of a dishonest majority (\( t < n \)), as the protocols provide stronger security guarantees against active adversaries. 
Even though honest majority protocols might seem appropriate given that all nodes are deployed at the \gls{fsp} level by a single entity, we choose 
dishonest majority due to stronger security guarantees against active adversaries. 
However, this comes at the cost of robustness, which cannot be achieved under dishonest majority settings \cite{cryptoeprint:2022/062,cryptoeprint:2014/668}.

\textbf{Computational Power:} Given the active security setting with a dishonest majority, the protocol should aim to achieve computational security as perfect security is unattainable under these settings \cite{cryptoeprint:2022/062}.

\textbf{Corruption Power:}
This model assumes static corruptions, as frequent key changes in Filia eliminate the need for proactive security or handling mobile adversaries. 
Key updates occur naturally during transactions, making explicit key refreshing after specific epochs unnecessary. 
However, this assumption can be extended to include proactive security measures for wallets that remain inactive for extended periods.

\subsubsection{Network and Communication Model}
Our system assumes all signatories are connected through an authenticated and synchronous broadcast mechanism within a single \gls{fsp}. 
This setup ensures that communication is public, and any message sent will either be received by all participants in the next round or not be delivered at all. 
Without pre-established authenticated communication, an adversary who previously controlled a corrupted party and managed the communication channels can impersonate that party indefinitely \cite{cryptoeprint:1998/012}.

The synchronous broadcast mechanism is crucial for maintaining accountability and ensuring all parties reach a consensus. 
Without defined communication delays, holding any signatory accountable for failing to respond would be impossible.
Additionally, reliable broadcast is vital for distributing proofs during the \gls{dkg} process and upholding accountability throughout the protocol \cite{cryptoeprint:2021/060}.

\subsection{Selection of Threshold \gls{ecdsa} Scheme}
\glspl{tss} have been extensively explored for their role in securing digital assets. 
Table~\ref{tab:libraries} compares some popular libraries available and implementations of \glspl{tss}.

Two comprehensive surveys \cite{cryptoeprint:2020/1390,sedghighadikolaei2024comprehensivesurveythresholdsignatures} provide an overview of key developments in the field. 
\gls{sota} protocols such as CGGMP21~\cite{cryptoeprint:2021/060} and DKLs23~\cite{cryptoeprint:2023/765} build on earlier advancements~\cite{cryptoeprint:2020/492,cryptoeprint:2020/540,cryptoeprint:2018/987,cryptoeprint:2018/499,cryptoeprint:2019/523}. 
These protocols follow standard steps:
\begin{itemize}
    \item Rewriting the ECDSA signing equation into an ``\gls{mpc}-friendly'' equivalent.
    \item Employing cryptographic primitives for secure multiplication (e.g., Oblivious Transfer, Paillier, or Class Groups).
    \item Verifying that all operations are performed honestly.
\end{itemize}
The primary distinctions between these protocols lie in their security definitions, underlying assumptions, and communication and computation complexities, which depend on the used cryptographic primitives. 
Table~\ref{tab:comparison} provides a summary of recent threshold signature schemes, detailing key differences in threshold settings, robustness, corruption strategies, and security assumptions. 

\begin{table}[t]
    \scriptsize\centering
    \begin{tabular}{p{2.5cm} c c p{6.5cm}}
        \toprule
        \textbf{Scheme} & \textbf{Robust} & \textbf{Corruption} & \textbf{Assumptions} \\
        &  & \textbf{Strategy}  &\\
        \midrule
        \multicolumn{4}{c}{Honest Majority} \\
        \midrule
        Gennaro et al. \cite{eurocrypt-1996-2312}   & \yes  & Static & None beyond unforgeability of DSS signatures \\
        Damg\r{a}rd et al. \cite{cryptoeprint:2020/501}  & \yes  & Static & None beyond ECDSA security \\
        Pettit \cite{cryptoeprint:2021/1386} & \yes  & Static & \gls{ddh}, ECDSA unforgeability  \\
        Bonte et al. \cite{cryptoeprint:2020/214} & \no  & Static & \gls{dlp}, MPC-friendly hash function \\
        Shi et al. \cite{10.1145/3545948.3545977}  & \yes  & Static & None beyond EdDSA security \\
        Stinson and Strobl \cite{Stinson2001ProvablySD} & \yes  & Static & \gls{dlp} \\
        Komlo and Goldberg \cite{cryptoeprint:2024/466} & \yes  & Static & \gls{dlp}, Secure hash function \\
        Kondi and Ravi \cite{KR24} & \no  & Static & \gls{pki}\\
        Jonathan Katz and Antoine Urban \cite{cryptoeprint:2024/2011} & \no  & Static & Semi-honest coordinator, unforgeability of standard ECDSA\\
        \midrule
        \multicolumn{4}{c}{Dishonest Majority} \\
        \midrule
        Canetti et al. \cite{cryptoeprint:2021/060} & \no  & Adaptive & Strong-RSA, ECDSA unforgeability, semantic security of Paillier encryption, \gls{ddh} \\
        Abram et al. \cite{cryptoeprint:2021/1587} & \no  & Static & Ring-LPN, ECDSA unforgeability \\
        Doerner et al. \cite{cryptoeprint:2023/765} & \no  & Static & Ideal commitment and two-party multiplication primitives \\
        Feng et al. \cite{cryptoeprint:2024/358} & \no & Adaptive & LPN, PCF \\
        Komlo and Goldberg \cite{cryptoeprint:2020/852} & \no  & Static & \gls{dlp} \\
        Ruffing et al. \cite{cryptoeprint:2022/550}  & \yes  & Adaptive & \gls{dlp} \\
        Lindell \cite{cryptoeprint:2022/374} & \no  & Static  & Ideal commitment and \gls{pki} \\
        \bottomrule
    \end{tabular}

    \caption[Comparison of Threshold ECDSA, EdDSA, and Schnorr Signature Schemes.]{Comparison of Threshold ECDSA, EdDSA, and Schnorr Signature Schemes. \yes~indicates that the property is achieved, while \no~indicates that it is not}
    \label{tab:comparison}
\end{table}

Among the libraries evaluated, the implementation of CGGMP21 by \emph{Lockness} \cite{dfns} meets our key system requirements. 
First, unlike other libraries, it is production code, not a \gls{poc} such as silence-labs~\cite{silence-laboratories}.
Next, the library supports secp256r1, the elliptic curve that Filia (\textbf{R6}) requires. 
Additionally, it enables essential functionalities such as importing existing keys (\textbf{R5}) and exporting keys securely from nodes (\textbf{R4}).
One of the key advantages of CGGMP21 is its efficient signing process, which is crucial for achieving high throughput in the Filia system. 
The protocol employs pre-signing and a one-round online signing process. 
This allows most cryptographic work to be completed ahead of time, minimizing delays during transactions and directly addressing the system’s need for high signature throughput (\textbf{R3}). 
CGGMP21 is a \gls{sota} threshold \gls{ecdsa} scheme with four signing rounds. 

The protocol’s security is further enhanced by its \gls{uc} \cite{cryptoeprint:2000/067}, ensuring that CGGMP21 remains secure even when composed with other cryptographic protocols. 
The \gls{uc}-security guarantee fulfills the requirement for strong security (\textbf{R7}). 
Thus, the combination of compatibility, efficiency, and security makes CGGMP21 an optimal choice for the Filia system, particularly with the \emph{Lockness} \cite{dfns} library.

\begin{table}[t]
    \scriptsize
    \centering
    \begin{tabular}{l c c c c c c}
    \toprule
    \textbf{Library} & \textbf{Curves} & \textbf{Key Export/} & \textbf{Audited} & \textbf{Lang.} & \textbf{Docs} & \textbf{Purpose} \\
        & & \textbf{Import} & &  &  &  \\
    \midrule
    \multicolumn{7}{c}{Honest Majority} \\
    \midrule
        Block-Daemon \cite{Blockdaemon}  & P-224/256/384/521, & \yes & \yes & Go, Java, Node & +++  & Prod. \\
         & Ed25519, Ed448 & & & &   &  \\ 
        Nakasendo \cite{nakasendo} &  & \no & \no & C++, Python & +  & Undefined \\ 
    \midrule
    \multicolumn{7}{c}{Dishonest Majority} \\
    \midrule
        Lockness \cite{dfns} & secp256r1, stark & \yes & \yes & Rust & ++  & Prod. \\ 
        circlefin \cite{circle} & ed25519 & \no & \no & Go & +  & Prod. \\ 
        silence-labs \cite{silence-laboratories} &  & \no & \no & Rust & +  & \gls{poc} \\ 
        0xCarbon \cite{0xCarbon} &  & \no & \no & Rust & +  & Undefined \\ 
        ZenGo-X \cite{ZenGo-X} &  & \no & \yes & Rust & ++  & \gls{poc} \\ 
        alephzero \cite{alephzero} &  & \no & \no & Go & -  & \gls{poc} \\ 
        taurushq-io \cite{taurushq-io} &  & \no & \yes & Go & +  & Undefined \\ 
        cait-sith \cite{cait-sith} &  & \no & \no & Rust & +  & \gls{poc} \\ 
        bnb-chain \cite{bnb-chain} & Edwards & \no & \yes & Go & ++  & Prod. \\ 
        Protect \cite{protect} &  & \no & \no & Java & +++  & \gls{poc} \\ 
        Safe-heron \cite{Safeheron} & P-256, stark & \no & \yes & C++ & +  & Prod. \\ 
    \midrule
    \multicolumn{7}{c}{Framework} \\
    \midrule
    MP-SPDZ \cite{mp-spdz} & P-256 & \yes & ? & Python, C++ & +++  & \gls{poc} \\ 
    \bottomrule
    \end{tabular}
    \caption{Comparison of software libraries of \glspl{tss}. All support at least secp256k1, so we do not list it separately. The \textit{Docs} column indicates the quality of documentation: more ``+'' or ``-''  for better or poorer documentation, respectively. \yes~indicates the feature is supported}
    \label{tab:libraries}
\end{table}

\section{Theoretical and Empirical Evaluations}
The section provides an overview of the experiment design and the evaluation findings.

\subsection{Security Analysis}
As a part of the theoretical evaluation, we consider the communication and computational complexities of the CGGMP21 protocol. 
This information provides details about the scalability of our solution, which is a critical consideration for its applicability.
Table~\ref{tab:comp-comm} summarizes the complexities from the perspective of a single node, including the respective communication rounds and Big(O) cost. 

The security of the \gls{kmn} falls back on the security of the CGGMP21 protocol, as implemented in the \textit{Lockness} library, which has been rigorously audited. 
However, additional considerations for \gls{kmn} security include the security of performing background \gls{dkg} and key assignments. 
The execution of \gls{dkg} in the background does not compromise overall security, as unauthorized access to the database of a node would expose the key shares of that node, regardless of when or how they are generated. 
Moreover, considering the system architecture, including the \gls{pp}, an adversary who gains control of the \gls{pp} would not be able to forge signatures. 
This is because every transaction request to the \gls{pp} requires an authentication token tied to the user, such as biometric verification stored securely on the user device enclave. 
While an attacker in this scenario could disrupt operations, such activity would be detectable, allowing for a prompt restart of the \gls{pp} to restore functionality. 
Notably, a coordinator is introduced in the implementation because it simplifies communication to facilitate the implementation of \gls{poc}.

The \textit{Lockness} library operates under two primary assumptions: (1) all messages are authenticated, and (2) all \gls{p2p} messages are encrypted. 
These requirements are fulfilled using the \textit{libp2p} library, which ensures secure and encrypted communication channels. 
Additionally, the library assumes reliable broadcast, achieved using an integrated reliability check mechanism. 
This mechanism enables each participant to hash the messages received in the previous round and include these hashes in their outgoing messages. 
By comparing these hashes, participants verify that all parties have received identical messages, ensuring consistency and reliability throughout the protocol execution.

\begin{table}[t]
   \centering
   \begin{tabular}{c c c c}
   \toprule
   \textbf{Phase} & \textbf{Round} & \textbf{Computation} & \textbf{Communication} \\
   \midrule
   \textbf{Key Generation}
       & Round 1 & $O(1)$ & $O(n)$ \\ 
       & Round 2 & $O(1)$ & $O(n)$ \\ 
       & Round 3 & $O(n)$ & $O(n)$ \\ 
       & Output & $O(n)$ & no communication\\ 
   \midrule
   \textbf{Presigning}
       & Round 1 & $O(t)$ & $O(t)$ \\ 
       & Round 2 & $O(t)$ & $O(t)$ \\ 
       & Round 3 & $O(t)$ & $O(t)$ \\ 
       & Output & $O(t)$ & no communication\\
   \midrule
  \textbf{Signing}
       & Round 1  & $O(1)$ & $O(t)$ \\ 
       & Output & $O(t)$ & no communication \\
   \bottomrule
   \end{tabular}
   \caption{Complexities for CGGMP21 protocol \cite{cryptoeprint:2021/060,glas2022evaluation}}
   \label{tab:comp-comm}
\end{table}

\subsection{Experiment Design \& Empirical Evaluation}

The implementation is based on the Rust-based \textit{Lockness} library~\cite{dfns}. 
The library abstracts the network layer, requiring only stream-based input and output for message exchange. 
For this, we rely on the \textit{libp2p} library~\cite{libp2p2023} to handle peer-to-peer communication.

We adopt the \gls{kmn} approach (i.e., \textbf{Solution 2}). 
In this setup, each \gls{fsp} integrates its \gls{kmn}. 
Keys and pre-signatures are generated in advance by the \gls{kmn} nodes. 
When a \gls{pp} sends a key generation request, the \gls{kmn} assigns an unallocated key to the request. 
Pre-signatures are utilized for signature operations on Filia commands; otherwise, the system falls back to an interactive signing step where the \gls{pp} waits for the \gls{kmn} nodes to interactively generate a signature. 
We introduce a coordinator node between the client \gls{pp} and the \gls{kmn} nodes. 
This node collects key shares from the \gls{kmn} nodes to reconstruct the full key during key export. 
It also manages key import operations by receiving a full key and splitting it into \( n \) key shares for the key addition/update process. 
Additionally, the coordinator node aggregates pre-signatures (also known as partial signatures) from the \gls{kmn} nodes to produce the final signature, which is then sent back to the client.

The evaluation is conducted in two phases. 
First, we benchmark cryptographic operations independently to analyze performance under varying  \( n \) and  \( t \). 
These should provide insights into the cost following the overview in Table~\ref{tab:comp-comm}.
Second, we integrated the updated protocol into the Filia system and measured \gls{e2e} transaction performance, including latency and throughput.

The cryptographic operations are evaluated in a local testbed using the \gls{methoda} \cite{Rezabek2022EngineJNSM,rezabek2023preprint} framework.
Our experiments run on four dedicated nodes. 
Each dedicated node has 64 \gls{vCPUs} and \SI{768}{\giga\byte} of RAM. 
Each experiment includes application code and service management files, which handle configuration and execution across nodes. 
Within \gls{methoda}, scenarios define sequences of experiments, allowing controlled adjustments to parameters like \(n\) and \(t\).
The nodes are interconnected over a dedicated test switch. 
To assess the scalability of the solution, we carry out experiments with various values of \(t\) and \(n\) from \( t = 2, n = 3 \) to \( t = 32, n = 52 \). 
We evaluate individual steps, including \gls{dkg}, pre-signing, and signing.
Each experiment involves 100 iterations and uses a single thread per node. 
These parameters are chosen based on evaluations of other threshold signature schemes (e.g., FROST \cite{cryptoeprint:2020/852}, GG18 \cite{cryptoeprint:2019/114}, and GG20 \cite{cryptoeprint:2020/540}) and relevance to expect the scale of the \gls{cbdc} deployments.
We fix the message size to \SI{256}{\bit}.
We measured the computation and I/O time for each step during the experiment run. 
The \textbf{computation time} is the time each node spends on local cryptographic operations.
On the other hand, the \textbf{I/O time} is the total time each node waits for messages to proceed to the next round.
Therefore, the total iteration time is the sum of computation and I/O times.
This evaluation uses the PerfProfiler utility from the \textit{Lockness} library.
Time measurement begins at the start of the iteration once the room ID is assigned and all nodes are ready to execute the protocol. 
This assumes all expected nodes can communicate with each other.

Each epoch begins when all nodes in the protocol discover and authenticate each other. 
Each node waits for a defined number of nodes \(n\) and then assigns each protocol iteration a unique room ID with a dedicated communication channel for that iteration's messages. 
Time measurement begins at the start of the iteration once the room ID is assigned and all nodes are ready to execute the protocol.

\subsubsection{Results of Phase 1}
\begin{figure}[t]
    \begin{subfigure}{.49\linewidth}
        \includegraphics[width=\linewidth]{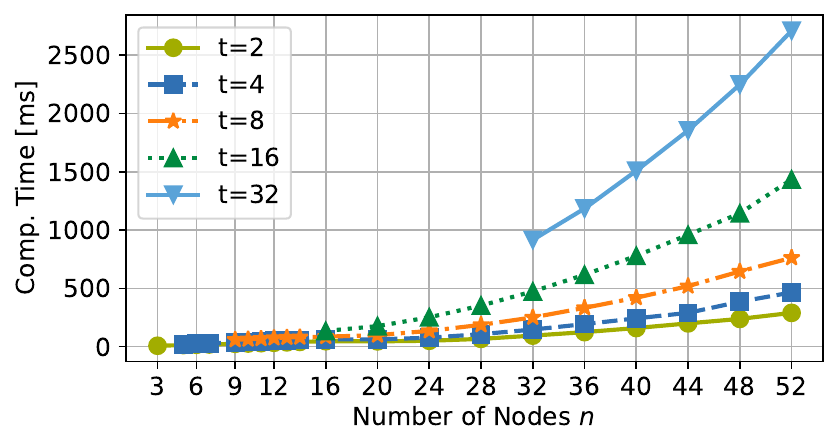}
        \caption{Computation Time}\label{fig:dkg-time}
    \end{subfigure}
    \hfill
    \begin{subfigure}{.49\linewidth}
        \includegraphics[width=\linewidth]{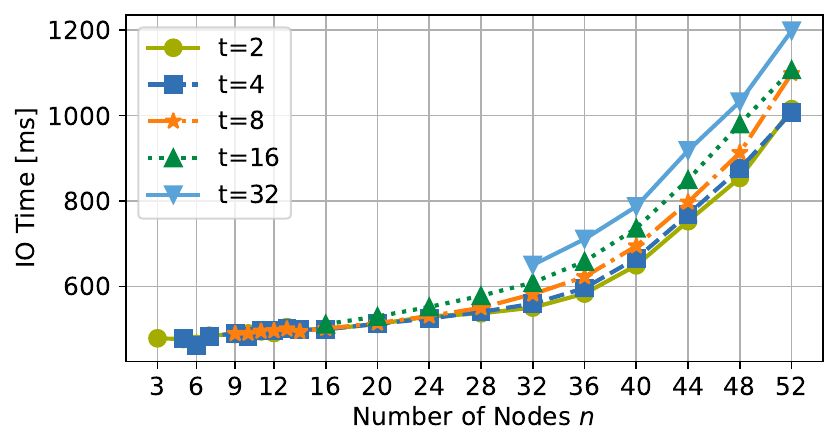}
        \caption{I/O Computation Time}\label{fig:dkg-io}
    \end{subfigure}
    \caption{\gls{dkg} Time vs Number of Nodes}\label{fig:dkg}
\end{figure}

First, we conduct experiments of the \gls{dkg} as shown in Figure~\ref{fig:dkg}.
The results show that increasing the number of nodes and the threshold directly impacts the time required for \gls{dkg}. 
As shown in Figure~\ref{fig:dkg-time}, the time increases quadratically with the number of nodes and threshold. 
This is also the case for the I/O time when each node has to wait to proceed, as shown in Figure~\ref{fig:dkg-io}.
We observe that for a smaller number of nodes, the I/O time exceeds computation time, likely due to communication delay among nodes. 
However, computation time begins to dominate the total \gls{dkg} duration as the number of nodes and threshold increase, observing computation time up to \SI{2500}{\milli\second} for $n=52$ and $t=32$. 
More time is spent on computation than on I/O. 

\begin{figure}[t]
    \begin{subfigure}{.49\linewidth}
        \includegraphics[width=\linewidth]{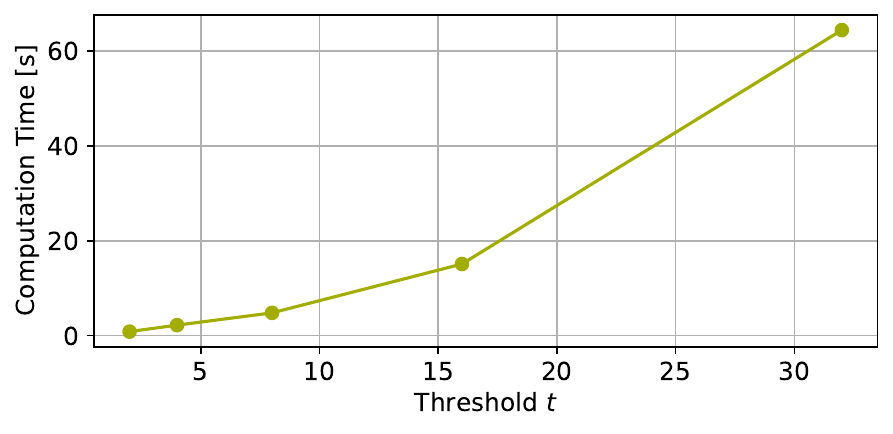}
        \caption{Computation Time}\label{fig:presign-time}
    \end{subfigure}
    \hfill
    \begin{subfigure}{.49\linewidth}
        \includegraphics[width=\linewidth]{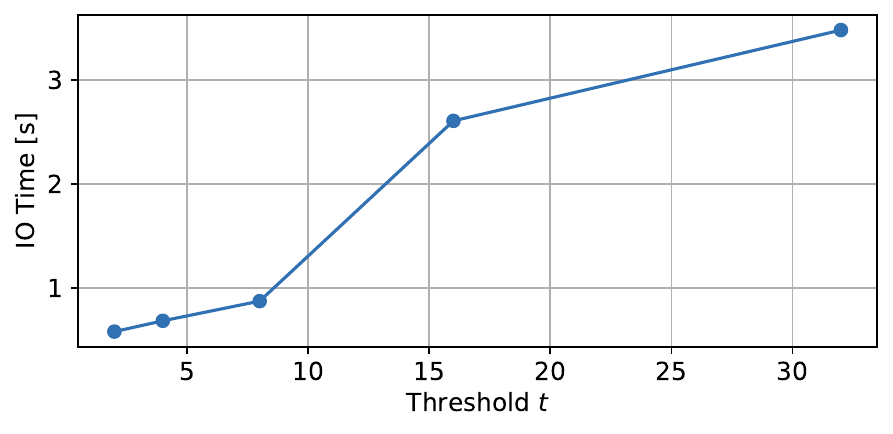}
        \caption{I/O Computation Time}\label{fig:presign-io}
    \end{subfigure}
    \caption{Pre-signing Time vs Threshold}\label{fig:pre-sign}
\end{figure}

\begin{figure}[t]
    \begin{subfigure}{.49\linewidth}
        \includegraphics[width=\linewidth]{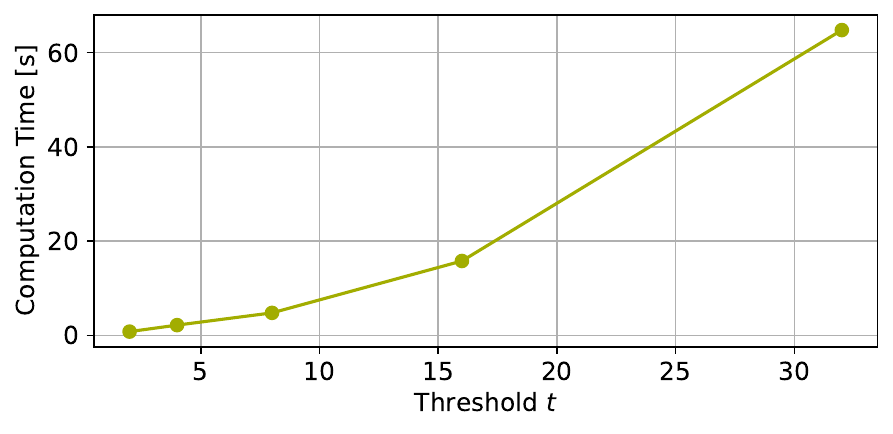}
        \caption{Computation Time}\label{fig:sign-time}
    \end{subfigure}
    \hfill
    \begin{subfigure}{.49\linewidth}
        \centering
        \includegraphics[width=\linewidth]{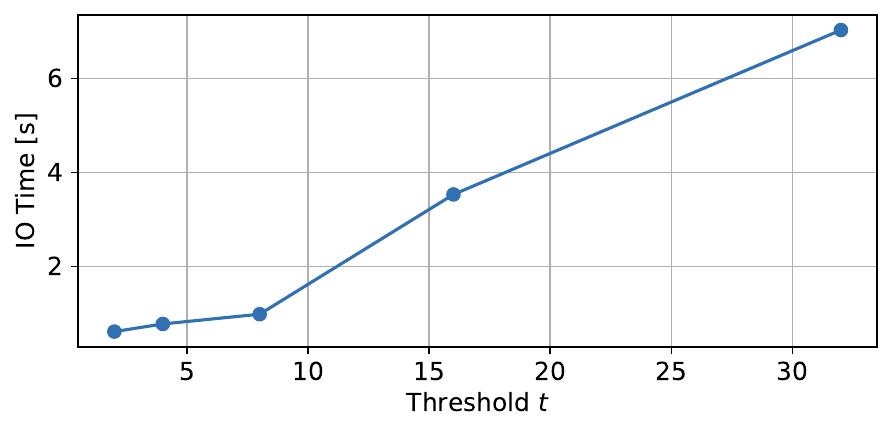}
        \caption{I/O Computation Time}\label{fig:sign-io}
    \end{subfigure}
    \caption{Signing Time vs Threshold}\label{fig:sign}
\end{figure}

As pre-signing and signing are not dependent on $n$, we average all values for respective computation and I/O times.  
As shown in Figure~\ref{fig:pre-sign}, the pre-signing follows a similar tendency for increasing $t$. 
Figure~\ref{fig:presign-time} shows computation time with a quadratic increase. 
For the I/O time, we observe fractional cost in comparison to the computation time, as shown in Figure~\ref{fig:presign-io}.
Continuing with signing presented in Figure~\ref{fig:sign} we observe some order of times as for pre-signing.
Especially the Figure~\ref{fig:sign-time} is almost identical to the Figure~\ref{fig:presign-time}.
However, for I/O time we observe an increase, as introduced in Figure~\ref{fig:sign-io}.
The time difference between pre-signing and signing is minimal, indicating that the final online round, which produces the signature on the known message, is computationally inexpensive.

\subsubsection{Results of Phase 2}
\begin{figure}[t]
    \begin{subfigure}{.49\linewidth}
        \centering
        \includegraphics[width=\linewidth]{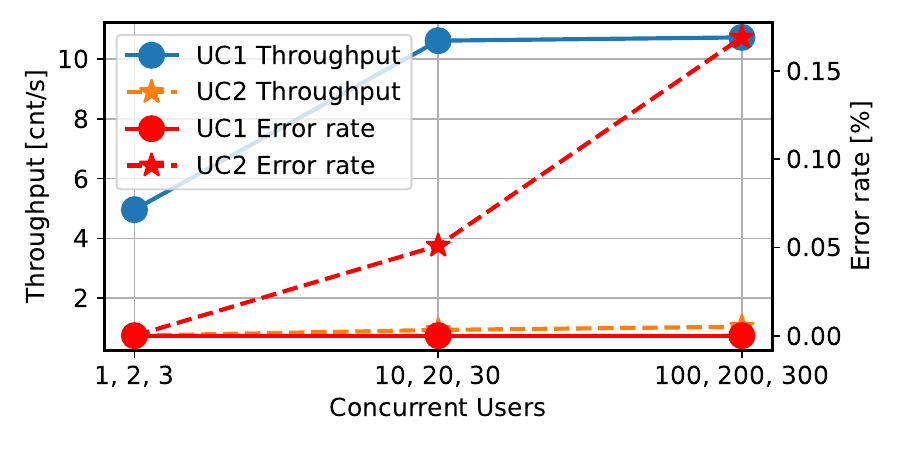}
        \caption{Total throughput w/ increasing concurrent users}\label{fig:througput}
    \end{subfigure}
    \hfill
    \begin{subfigure}{.49\linewidth}
        \centering
        \includegraphics[width=\linewidth]{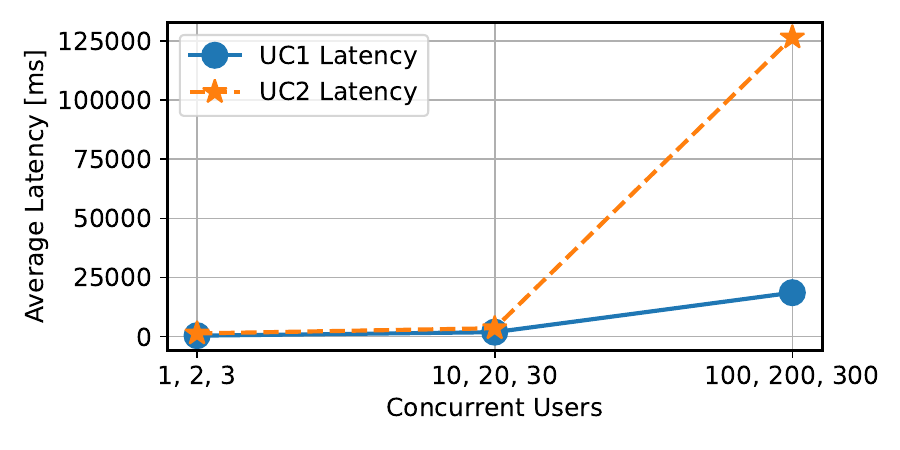}
        \caption{E2E latency with increasing concurrent users}\label{fig:latency}
    \end{subfigure}
    \caption{E2E transactions latency and throughput}\label{fig:latency-throughput}
\end{figure}

The integration testing is carried out locally on a MacBook M3 Pro due to the complexity of setting up a private cloud-controlled environment and the challenges of porting the code to external infrastructure. 
The setup includes two \glspl{kmn} (\( t = 2, n = 3 \)), two \glspl{fsp} hosting \glspl{pp}, and a central Verifier hosted in a Docker container to validate transactions. 
We test two \glspl{ucs}: \gls{ucs}1 transfers between customers of the same \gls{fsp}, whereas \gls{ucs}2 emulates transfers between different \glspl{fsp}. 
The evaluation simulated varying loads with 1, 2, 3, 10, 20, 30, 100, 200, and 300 concurrent users over 30-minute test sessions. 
Wallets were pre-created using APIs exposed by the \glspl{pp}, with each wallet assigned sufficient funds to execute transfers. 
The load tests measured \gls{e2e} transaction timing, starting from the API call to initiate the transfer until the recipient's wallet confirmed transaction completion.
As shown in Figure~\ref{fig:througput}, throughput increases steadily as the load increases in \gls{ucs}1, while in \gls{ucs}2, throughput improves slightly. 
In both cases, the latency increases as the load increases, as shown in Figures~\ref{fig:througput} and~\ref{fig:latency}. 
Performance evaluations reveal an order-of-magnitude decrease in throughput when using threshold signatures, particularly for cross-\gls{fsp} transfers. 
Under high loads (100, 200, and 300 concurrent users), \gls{ucs}2 throughput drops to approximately \SI{1.2}{} \gls{tps}, compared to \SI{60}{} \gls{tps} in a non-threshold configuration. 
Similarly, \gls{ucs}1 throughput decreases from 190 to around 10.7 \gls{tps} in the same setup.
We also observe failed transactions with more concurrent users for throughput with an error rate of up to \SI{0.15}{\percent}, as shown in Figure~\ref{fig:througput}. 

\subsection{Discussion}

The CGGMP21 protocol and the \gls{kmn} architecture collectively address all the requirements described, making them a fitting choice for Filia and other \gls{cbdc} systems. 
However, further research is necessary to address performance challenges with threshold cryptography.
As the number of nodes and the threshold increase, the performance of the cryptographic operations degrades noticeably. 
This is due to the growing computational and communication complexity, which scales with \(n\) and \(t\), as detailed in Table~\ref{tab:comp-comm}. 
In practical deployments, a smaller number of nodes is often sufficient to maintain better performance, and unless operated properly, larger node counts do not necessarily yield more security advantages.
It is important to note that integration testing results are based on executions in a testing environment; a speedup is expected in a production-ready setup, provided the environment is carefully tuned to optimize performance and minimize resource constraints.
While the added latency of communication between the \gls{pp} and the \gls{kmn} introduces a slight overhead, the system's scalability is mainly influenced by bottlenecks identified during the \gls{poc}. 

A key bottleneck appears in the time spent during signing operations without a pre-signature. 
This delay is introduced during the key import and update step, where the new key \( r_1 + r_2 \)  has no pre-signature (\encircled{7} in Figure~\ref{fig:transaction}). 
Consequently, a complete interactive signing on the \gls{kmn} side is required when merging and signing with this key. 
Sometimes, it is triggered by an interactive signing or key generation. 
This happens when the pre-generated keys are exhausted, or a key exists without a pre-signature. 
Therefore, the nodes have to re-initialize communication and authentication with each other, leading to higher latency and resource inefficiencies. 
Addressing these bottlenecks may involve triggering a pre-signing operation in the background upon key updates.
Similarly, nodes can establish communication channels among different protocol executions to reduce the overhead of re-establishing the secure channel between nodes.

The results for \gls{uc}2 further indicate a non-zero error rate, with transaction timeouts caused by resource constraints. 
Transactions exceeding a 5-minute processing time are classified as errors, often due to unhandled temporary issues or resource locking by other jobs. Notably, an error rate is also observed in non-threshold transactions, indicating that the issue is not solely attributable to the use of \gls{tss} but rather to resource availability and system load. A potential solution to mitigate these errors would be horizontal scaling, allowing the system to dynamically allocate additional resources as transaction loads increase.
In contrast, \gls{uc}1 demonstrates zero errors, reflecting the lower complexity of same-\gls{fsp} transfers compared to cross-\gls{fsp} transfers. 

The experimental design of this evaluation follows a logical progression that validates the solution from \gls{poc} to performance assessment. 
The \gls{e2e} transaction evaluations reveal the impact of \glspl{tss} on performance under various load conditions, replicating real-world transaction demands. 
Isolating cryptographic operations allows a detailed examination of resource costs associated with \gls{dkg}, pre-signing, and signing. 
Although \gls{dkg} and pre-signing occur as background tasks, their isolated evaluation highlights the computational demands they introduce. 
The signing evaluation shows the effect of the key update bottleneck, which requires interactive signing at that step. 
A configuration of \(t=2\) and \(n=3\) requires approximately one second for an interactive signing process.
While manageable under a small load, this latency represents a substantial performance cost as the system scales.
\section{Conclusion}

Our work improves the security and resilience of \gls{cbdc} systems, using Filia as a representative platform to address typical requirements. 
A comprehensive theoretical analysis identifies key system requirements, including compatibility with existing infrastructures, high signature throughput, secure key reconstruction, key updates, and strong security guarantees. 
We also define the computational and communication complexity of the CGGMP21 protocol and establish a robust security model to address adversarial behaviors.
Building on this foundation, the CGGMP21 protocol is selected and used to build a \gls{poc}, along with the design and implementation of a \gls{kmn} architecture. 
Empirical evaluation shows that the evaluation of cryptographic operations aligns with the complexity analysis. 
Moreover, \gls{e2e} transaction performance decreases by an order of magnitude compared to non-threshold systems, particularly in cross-\gls{fsp} transactions. 
Despite this, the system demonstrates acceptable performance under varying loads for smaller values of $n$, confirming the feasibility of \gls{tss} usage in practical \gls{cbdc} deployments like Filia while addressing key security challenges.

\subsubsection{Future Work}
As outlined, we identified several parts for optimization, which aim to improve the computation and communication bottlenecks. 
Similarly, even though the dishonest majority setting fits the \gls{cbdc} setting, investigating further \gls{tss} tailored to the honest majority setting may offer reduced complexity and improved efficiency for \gls{kmn}. 
Last, since \gls{cbdc} solutions will be used for the long term, investigating post-quantum secure \glspl{tss} is essential for their future-proofness.
Further efforts can focus on optimizing the computational and communication overhead of threshold operations, particularly addressing the identified bottlenecks.

\begin{credits}
\subsubsection{Acknowledgments}
We thank Franziska Kreitmair for early discussions on this subject and the anonymous reviewers for suggesting improvements to this paper.
This work has been partially supported by the German Federal Ministry of Research, Technology and Space (BMFTR), Verbundprojekt CONTAIN (13N16582) and POST (16KIS2159); the Bavarian Ministry of Economic Affairs, Regional Development, and Energy under project 6G Future Lab Bavaria; and Horizon Europe under project SLICES-PP (10107977).
\end{credits}

\printbibliography
\end{document}